\documentclass[aps,twocolumn, showpacs]{revtex4}
\usepackage{graphicx}
\usepackage{amsfonts}
\usepackage{amssymb,amsmath}

\begin{document}

\title{Distance between quantum states  in  presence of initial qubit--environment correlations: a comparative study}
 \author{Jerzy  Dajka, Jerzy \L uczka}
\affiliation{Institute of Physics, University of Silesia, 40-007 Katowice, Poland }
 \author{Peter H\"{a}nggi}
\affiliation{Institute of Physics, University of Augsburg, D-86135 Augsburg, Germany}

\begin{abstract}
The time evolution of the {\it trace distance} between two  states  of  an open quantum system may increase due to initial  system-environment correlations, thus exhibiting a breakdown of distance contractivity of  the reduced dynamics. We analyze how the time evolution of the distance depends on the chosen distance measure. Here we elucidate  the behavior of the trace distance, the Hilbert-Schmidt distance, the Bures distance, the Hellinger distance and the quantum Jensen-Shannon divergence for two system-environment setups, namely a qubit bi-linearly coupled to an infinite and a finite size environment with the latter composed of harmonic oscillators.
\end{abstract}


\maketitle
%
\section{ Introduction}

Quantum states represented by density matrices $\rho$ can be determined   by quantum state tomography and compared using  various quantifiers.
Distances and other similarity measures provide a quantitative method to evaluate how close two
states are together or how  precisely a quantum channel can   transmit information.
Unfortunately, there is no  single, ideal measure of distinguishability of different states.   There are no criteria for the distance measure to be 'better than another'.    Even the  natural  requirement that the distance between states should have properties of a metric (i.e. identity of indiscernibles, symmetry and the triangle inequality) is relaxed in a case of {\it fidelity} which is  a celebrated statistical similarity measure.
Loosely speaking, two states are close to each other if the distance is small.
We also expect that two different distances are equivalent if  any two states
that  become closer to one another in the sense of one distance measure  become  also closer in the sense of the second, and vice versa.

There are diverse ways of introducing a notion of distance between two quantum states \cite{gil}.
Examples of such distance measures comprise the  trace distance, Hilbert-Schmidt distance,  Bures distance, Hellinger distance  and Jensen-Shannon divergence, to mention a few,  see  also in Refs. \cite{gil,luo,majtey,dodonov,nielsen}. These metrics possess distinct properties like being Riemannian, monotone (contractive), with bounds and relations among them \cite{karol}.

Let us recall that any positive and trace-preserving map $\cal E$ defined on the whole space of operators $\rho$ on the Hilbert space   is contractive with respect to a given distance $D[\rho_1, \rho_2]$  if
\begin{eqnarray}\label{contr}
D[{\cal E}(\rho_1),  {\cal E}(\rho_2) ] \le D[\rho_1, \rho_2].
\end{eqnarray}
In particular, when ${\cal E} = {\cal E}_t$ is a completely positive quantum dynamical semigroup  such that $\rho(t) = {\cal E}_t \rho(0)$, then contractivity means that
\begin{eqnarray}\label{contr1}
D[\rho_1(t) , \rho_2(t)  ] \le D[\rho_1(s), \rho_2(s)] \;,  \quad \mbox{for} \quad t > s.
\end{eqnarray}
As a consequence, the distance cannot increase in time and the  distinguishability of any states  can not increase above an initial value.   In particular, if a quantum  open system and its environment are initially prepared in an uncorrelated  state, the reduced dynamics is completely positive and hence contractive with respect to some metrics.
In consequence, the distance $D[\rho_1, \rho_2]$ between two states can tend to zero  when the system approaches a unique steady-state, i.e.  the dynamics is relaxing.

We emphasize that contractivity is not a universal feature but depends on the metric:  quantum evolution may be contractive with respect to a given metric and   may not be contractive with respect to other metric measures. Moreover,  contractivity of quantum evolution can break down provided that the system  is initially correlated with its environment.  Effects induced by such correlations have been  studied in various context \cite{pech,romero2004,correla,ban}. First experiments on initial system-environment correlations are reported in Ref. \cite{exper}.
Examples of an exact reduced dynamics which fail contractivity with respect to the {\it trace distance} are presented in Refs. \cite{daj,breu3}: the trace distance of different states  grows above its initial value and the distinguishability growth occurs not only at the short time scales but is shown  to be a feature of  the long-time limit as well. The trace metric is likely the most important measure of the size for distance in quantum information processing and  
according to Ref. \cite{breu3}, an increase of the distance can be interpreted
in terms of the exchange of information between the system and its environment. If the distance
increases over its initial value, information which is locally inaccessible at the initial time is transferred to the open system. This transfer of information enlarges the distinguishability of the open-system states which suggests various ways for the  experimental detection of initial correlations. 
With this study we  demonstrate that the correlation-induced distinguishability growth is not generic with respect to  distance measures but distinctly depends on the assumed form of the metric measure. 

The outline  of the paper is as follows.  In Sec. II  we  list several forms of the distance measure. In Sec. III, we define a dephasing model  of the qubit plus environment \cite{defaz} and environment is assumed to be infinite. We also present the reduced dynamics of the qubit  for a particular initial qubit-environment state which is correlated (entangled). Properties of time evolution of the distance 
between  two states of the qubit are demonstrated for selected metrics. In  Sec. IV, we consider the similar model but now  with a finite size  environment consisting of just one  boson.  We study  distances between two states and analyze its properties. Finally,  Sec. V  provides our summary and some conclusions.

\section{A selection of different distance measures}
The question of similarity between quantum states can have very different meanings depending on the context in which the question is posed. One can distinguish at least two main classes of problems. The first  is related to the geometric structure of a set of states, and the second is related to the statistical content of quantum states. These two classes are not disjoint due to the richness of links joining different quantifiers \cite{karol}. Here we limit our consideration to measures which are, or are expected (as the Jensen--Shannon divergence discussed below) to be a {\it metric}.
We will consider the following  types of the distance
between  any two states $\rho_1$ and $\rho_2$: \\
1. The use of the trace distance, i.e., 
\begin{eqnarray}\label{trace}
D_T[\rho_1, \rho_2]=\frac{1}{2}\mbox{Tr} \sqrt{(\rho_1-\rho_2)^2},
\end{eqnarray}
presents a contraction  in the sense discussed in the Introduction and is limited to the unit interval,
\begin{eqnarray}
0\le D_T[\rho_1, \rho_2] \le 1.  \nonumber
\end{eqnarray}
 The trace distance, being Euclidean, has  apart from it geometric characteristics, also a profound statistical meaning as a quantifier for 'statistical distinguishability' of quantum states \cite{nielsen}. Due to its universal character  the trace distance has been considered in the context of contractivity break down caused by the system--environment correlations \cite{daj, breu3}. In this paper it will serve as a natural reference for other measures to be compared with.

2. The space of density matrices describing states of a quantum system can be equipped with a very natural scalar product \cite{nielsen} leading to
 the Hilbert-Schmidt distance:
\begin{eqnarray}\label{hilbert}
D_{HS}[\rho_1, \rho_2]= \sqrt{\mbox{Tr} (\rho_1-\rho_2)^2}.
\end{eqnarray}
This distance is restricted by the inequality relation
\begin{eqnarray}
0\le D_{HS}[\rho_1, \rho_2] \le 2  D_T[\rho_1, \rho_2].  \nonumber
\end{eqnarray}
The Hilbert--Schmidt distance is of  Riemann type. Unfortunately it generally does not possess  the 'contractivity property' discussed in the Introduction. Fortunately enough, however,  archetype quantum systems such as  qubits constitute useful  exceptions, as  it will be discussed in further detail below.

3. There is a very elegant and deep geometric structure  useful for studying  general quantum systems, namely the Hilbert--Schmidt fibre bundle  \cite{karol}.  Its base manifold is equipped with a  natural metric \cite{karol}, i.e.  the  Bures distance,
\begin{eqnarray}\label{bures}
D_B^2[\rho_1, \rho_2]= 2 \left[1- \sqrt{F(\rho_1, \rho_2)}\right].
\end{eqnarray}
The Bures distance is contractive and can be expressed by the  fidelity
\begin{eqnarray}\label{fidelity1}
F(\rho_1, \rho_2) = \left[\mbox{Tr}\sqrt{\sqrt{\rho_1}\; \rho_2 \;\sqrt{\rho_1}} \right]^2
\end{eqnarray}
and hence, additionally to its geometric character,  the Bures distance inherits a clear statistical interpretation.  In this case
\begin{eqnarray}
0\le D_B[\rho_1, \rho_2] \le \sqrt{2}.  \nonumber
\end{eqnarray}
4. Among the variety of  distances between states there are measures whose definition  originated from the statistical interpretation of quantum states \cite{karol}. One of them is the so called
 Hellinger distance; i.e., 
 
\begin{eqnarray}\label{hellinger}
D_H^2[\rho_1, \rho_2]= \mbox{Tr} \left(\sqrt{\rho_1} - \sqrt{\rho_2}\right)^2 \nonumber\\
= 2\left[1- A(\rho_1, \rho_2)\right],
\end{eqnarray}
where the quantum affinity reads
\begin{eqnarray}\label{affinity}
A(\rho_1, \rho_2) =\mbox{Tr}\left(\sqrt{\rho_1}\; \sqrt{\rho_2}\right).
\end{eqnarray}
The Hellinger distance assumes values from the interval
\begin{eqnarray}
0\le D_H[\rho_1, \rho_2] \le \sqrt{2}.\nonumber
\end{eqnarray}
5. The notion of (information)-entropy occurs in almost all branches of physics as a tool of quantifying information or relative information contained in  states, either classical or quantum. There are certain technical difficulties in using certain types of information entropies \cite{lamb}. These measures are, in general, not metrics. The Jensen-Shannon divergence is a tool which allows one to overcome this sort of problem. It is defined in terms of a symmetrized relative entropy between states;  here, however, we use instead the following expression \cite{majtey}:
\begin{eqnarray}\label{JS}
D_{JS}^2[\rho_1, \rho_2]= H_N\left(\frac{\rho_1 + \rho_2}{2}\right)
-\frac{1}{2} H_N\left(\rho_1\right) -\frac{1}{2} H_N\left(\rho_2\right), \nonumber
\\
\end{eqnarray}
where
\begin{eqnarray}\label{von}
H_N\left(\rho\right) =- \mbox{Tr} [\rho \ln \rho]\nonumber
\end{eqnarray}
is the von Neumann entropy. This quantity takes values from the unit interval,
\begin{eqnarray}
0\le D_{JS}[\rho_1, \rho_2] \le 1.
\end{eqnarray}
Whether the Jensen-Shannon divergence is a metric for all mixed states remains an unsolved problem
\cite{lamb,briet}.

Below, we will consider one-qubit system (with a $N=2$ dimensional Hilbert space) for which
one can  represent the density matrices  in the form
\begin{eqnarray}\label{dens}
\rho_i = \frac{1}{2} [1+ \vec{r}_i\cdot \vec{\sigma}],    \quad i=1, 2,
\end{eqnarray}
where $\vec{r}_i =[x_i, y_i, z_i]$ is the Bloch vector and $\vec{\sigma} =[\sigma_x, \sigma_y, \sigma_z]$ are the Pauli matrices.
In this case, the  trace and Hilbert-Schmidt distances are equivalent,
 namely \cite{wang},
\begin{eqnarray}\label{trace1}
D_{HS}[\rho_1, \rho_2]= \sqrt{2} D_{T}[\rho_1, \rho_2].
\end{eqnarray}
This  distance is equal to the ordinary Euclidean distance
between the two states on the Bloch sphere, i.e.
$D_{HS}(\rho_1, \rho_2) = |\vec{r}_1 -   \vec{r}_2|$.
Moreover,  the expression for the Bures distance simplifies because  the fidelity assumes the form \cite{karol}
\begin{eqnarray}\label{fidelity}
F(\rho_1, \rho_2) = \mbox{Tr} (\rho_1 \rho_2)
 +2\sqrt{\mbox{det} \rho_1 \mbox{det} \rho_2 }.
\end{eqnarray}
The Helinger distance can explicitly be calculated using the relation for the affinity
(\ref{affinity}).
Then
the affinity is expressed by the relation \cite{luo}
\begin{eqnarray}\label{aff}
A(\rho_1, \rho_2) =\frac{\left(1+\sqrt{1-r_1^2} \right) \left(1+\sqrt{1-r_2^2}\right) +\vec{r_1} \cdot \vec{r_2}}
{\left(\sqrt{1+r_1} + \sqrt{1-r_1}\right) \left(\sqrt{1+r_2} +  \sqrt{1-r_2}\right)},  \nonumber \\
\ \ \
\end{eqnarray}
where $r_i^2= x_i^2 + y_i^2 + z_i^2$.
The Jensen-Shannon divergence  (\ref{JS}) is expressed by the von Neumann entropy which is given by
\begin{eqnarray}\label{entr}
H_N\left(\rho_i\right) = \ln 2 - \frac{1}{2} \ln(1-r_i^2) -\frac{r_i}{2} \ln \frac{1 +r_i}{1-r_i}.
\end{eqnarray}
It has been proved that for qubits the Jensen-Shannon divergence is a metric \cite{briet}.

In prior works \cite{daj,breu3},  examples showing that the trace distance of different states  can grow above its initial value have been presented.   Our objective here is to investigate whether  the growth of distance measure  is preserved as well  for the other metric measures introduced above.

 \section{Model A: Qubit coupled to infinite environment of oscillators}

In this section, we consider the same model as in  Ref. \cite{daj}. For the readers convenience and to keep the paper self-contained, we provide all necessary definitions and notation. The model consists of  a qubit $Q$ (two-level system)   coupled to its  environment $B$ and we limit our considerations to the case when the process of energy dissipation  is negligible and only pure dephasing 
 is acting as the mechanism responsible for decoherence of the qubit dynamics \cite{defaz}.
Such a system can  be described  by the Hamiltonian  (with $\hbar=1$)
\begin{eqnarray}\label{ham}
H= H_Q \otimes  {\mathbb{I}}_B +{\mathbb{I}}_Q  \otimes H_B + S^z \otimes H_I, \\
H_Q = \varepsilon S^z, \quad
H_B=\int_0^\infty d\omega \, h( \omega) a^\dagger(\omega)a(\omega), \\
H_I=\int_0^\infty d\omega  \left[ g^*(\omega)a(\omega) +g(\omega)a^\dagger(\omega)\right],
\end{eqnarray}
where  $S^z$  is the z-component of the spin operator and is represented by the diagonal matrix $S^z =diag[1,-1]$ of elements $1$ and $-1$. The parameter  $ \varepsilon$ is the  qubit energy splitting,   ${\mathbb{I}}_Q$   and ${ \mathbb{I}}_B$  are   identity operators (matrices) in corresponding Hilbert spaces of the  qubit $Q$ and the environment $B$, respectively.
The operators $ a^\dagger(\omega)$ and $a(\omega)$ are the bosonic creation and annihilation operators, respectively.  The real-valued spectrum function $h(\omega)$  characterizes the environment. The coupling is described  by the function $g(\omega)$ and  the function
 $ g^*(\omega)$ is   the complex conjugate to   $g(\omega)$.
The Hamiltonian (\ref{ham}) can be rewritten in the  block--diagonal structure \cite{fidel},
\begin{eqnarray} \label{H1}
H=diag[H_{+}, H_{-}],  \quad
H_\pm =  H_B  \pm   H_I  \pm  \varepsilon { \mathbb{I}}_B.
\end{eqnarray}
As an example, we assume a correlated  initial state of the total system in the form  similar to that in  Ref.  \cite{daj}, namely,
\begin{eqnarray}\label{ini}
|\Psi(0)\rangle =b_+|1\rangle \otimes |\Omega_0\rangle +b_-|-1\rangle \otimes
|\Omega_{\lambda}\rangle.
\end{eqnarray}
The states $|1\rangle$ and $|-1\rangle$ denote the excited and
ground state of the qubit, respectively. The non-zero complex numbers $b_+$ and $b_-$ are chosen  such that
$|b_+|^2 + |b_-|^2 =1$.  The state
 $|\Omega_0\rangle$   is  the  ground  (vacuum) state of the environment and
\begin{eqnarray}\label{omega1}
|\Omega_{\lambda} \rangle = C_{\lambda}^{-1} \, \left[(1-\lambda)|\Omega_0\rangle +\lambda |\Omega_f\rangle  \right],
\end{eqnarray}
where  $|\Omega_f\rangle =D(f) |\Omega_0\rangle$
is  the coherent state. The  displacement (Weyl) operator $D(f) $  reads \cite{brat}
\begin{eqnarray}\label{displacement}
D(f)=\exp\left\{\int_0^\infty d\omega \left[ f(\omega)a^{\dagger}(\omega) -  f^*(\omega)a(\omega)\right]\right\}
\end{eqnarray}
 for an arbitrary square--integrable function $f$.
The constant  $C_{\lambda}$ normalizes the state (\ref{ini}) and is given by the expression
\begin{eqnarray}\label{C}
C_{\lambda}^2=(1-\lambda)^2+\lambda^2+2\lambda(1-\lambda)
Re \langle \Omega_0|\Omega_f\rangle,
\end{eqnarray}
where ${Re}$ is a real part of the scalar product $\langle \Omega_0|\Omega_f\rangle $ of two states in the environment Hilbert space.
The  parameter $\lambda\in[0,1]$  controls  the   initial entanglement of the qubit with  environment.  For $\lambda=0$ the qubit  and the environment are initially uncorrelated while for  $\lambda=1$ the entanglement is  most prominent   for a given class of initial states.

The initial state  (\ref{ini})  of the total  system evolves according to the formula
\begin{eqnarray}  \label{ewol}
|\Psi(t)\rangle =b_+|1\rangle \otimes |\psi_+(t)\rangle +b_-|-1\rangle \otimes
|\psi_-(t)\rangle,
\end{eqnarray}
where
\begin{eqnarray}  \label{ewolB}
|\psi_+(t)\rangle &=& \exp(-i H_+ t)   |\Omega_0\rangle, \nonumber\\
|\psi_-(t)\rangle &=& \exp(-i H_- t)   |\Omega_{\lambda}\rangle.
\end{eqnarray}
 The density  matrix of the total (isolated) system is 
 $\varrho(t) = |\Psi(t) \rangle \langle\Psi(t)|$. In turn, the partial trace  $\mbox{Tr}_B$ over the environment $B$ yields
 the density  matrix  $\rho_{\lambda}(t) =  \mbox{Tr}_B \varrho(t)$ of the qubit.
It can be expressed in the matrix form as:
\begin{eqnarray}\label{ro}
\rho_{\lambda}(t)=\left(\begin{array}{cc} |b_+|^2 & b_+b_-^* \, A_{\lambda}(t) \\
b_+^*b_- \,   A_{\lambda}^*(t) & |b_-|^2 \end{array}  \right),
\end{eqnarray}
where the dephasing function $A_{\lambda}(t)$ reads
\begin{eqnarray}\label{A}
 A_{\lambda}(t)= C_{\lambda}^{-1} \,e^{-2i\varepsilon t - r(t)}  \left[1-\lambda+\lambda
e^{-2i\Phi(t) +  s(t)}  \right],
\end{eqnarray}
and  \cite{fidel}
\begin{eqnarray}\label{r}
r(t) &=& 4\int_0^\infty d\omega g_h^2(\omega) \left[1-\cos(\omega t) \right], \quad \quad\quad \quad\quad\nonumber\\
s(t) &=& 2\int_0^\infty d\omega g_h(\omega)f(\omega)\left[1-\cos(\omega t)\right]   \quad \nonumber\\
&-&  \frac{1}{2} \int_0^\infty d\omega f^2(\omega)
\end{eqnarray}
%
where $g_h(\omega) = g(\omega)/h(\omega)$ and
\begin{eqnarray}\label{Fi}
\Phi(t) =  \int_0^\infty d\omega g_h(\omega)f(\omega) \sin(\omega t).
\end{eqnarray}
Without loss of  generality we have assumed here that the  functions $g(\omega)$ and $f(\omega)$ are real valued.

\begin{figure}[htpb]
\includegraphics[width=0.2\textwidth, angle=270]{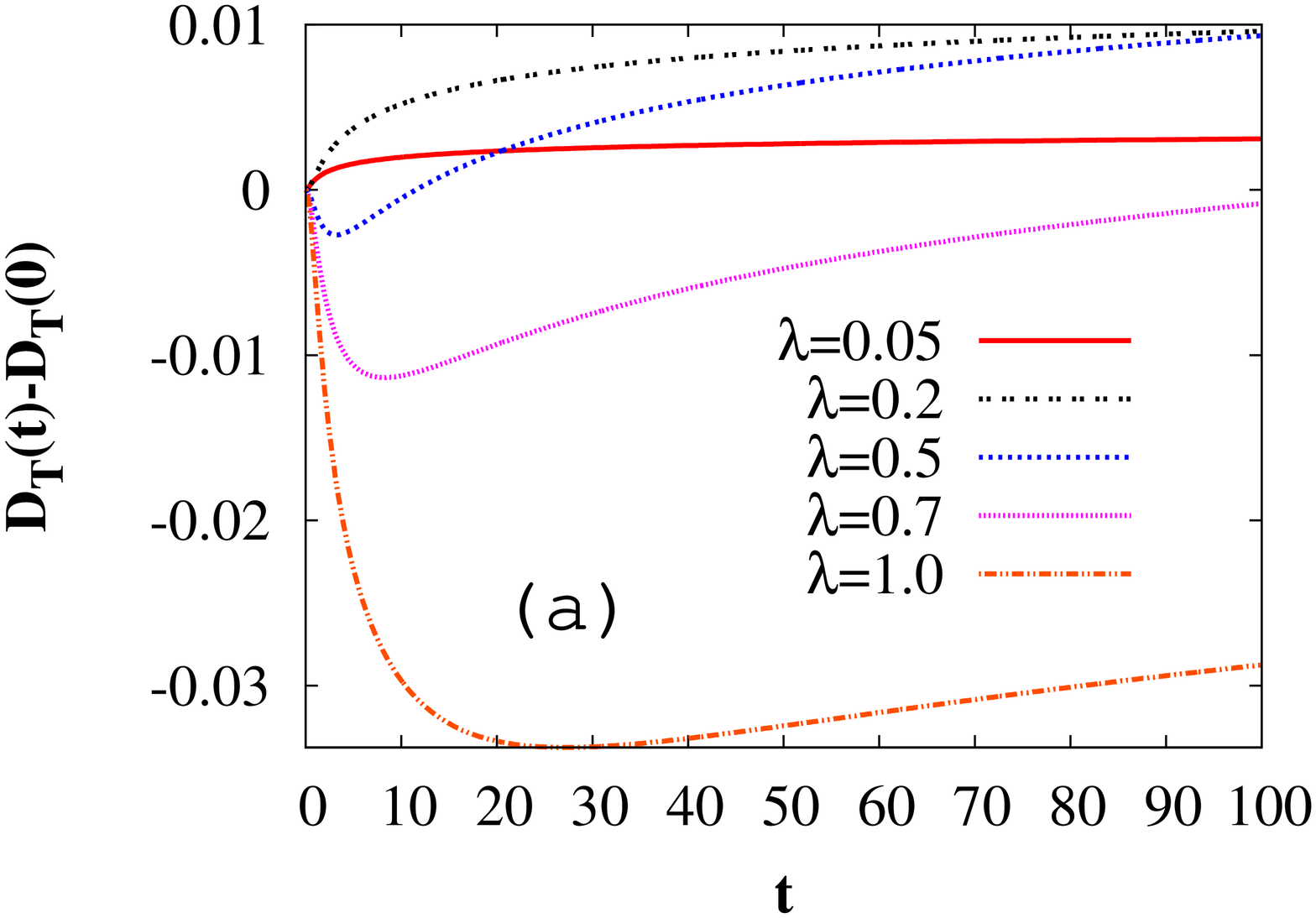}
\includegraphics[width=0.2\textwidth, angle=270]{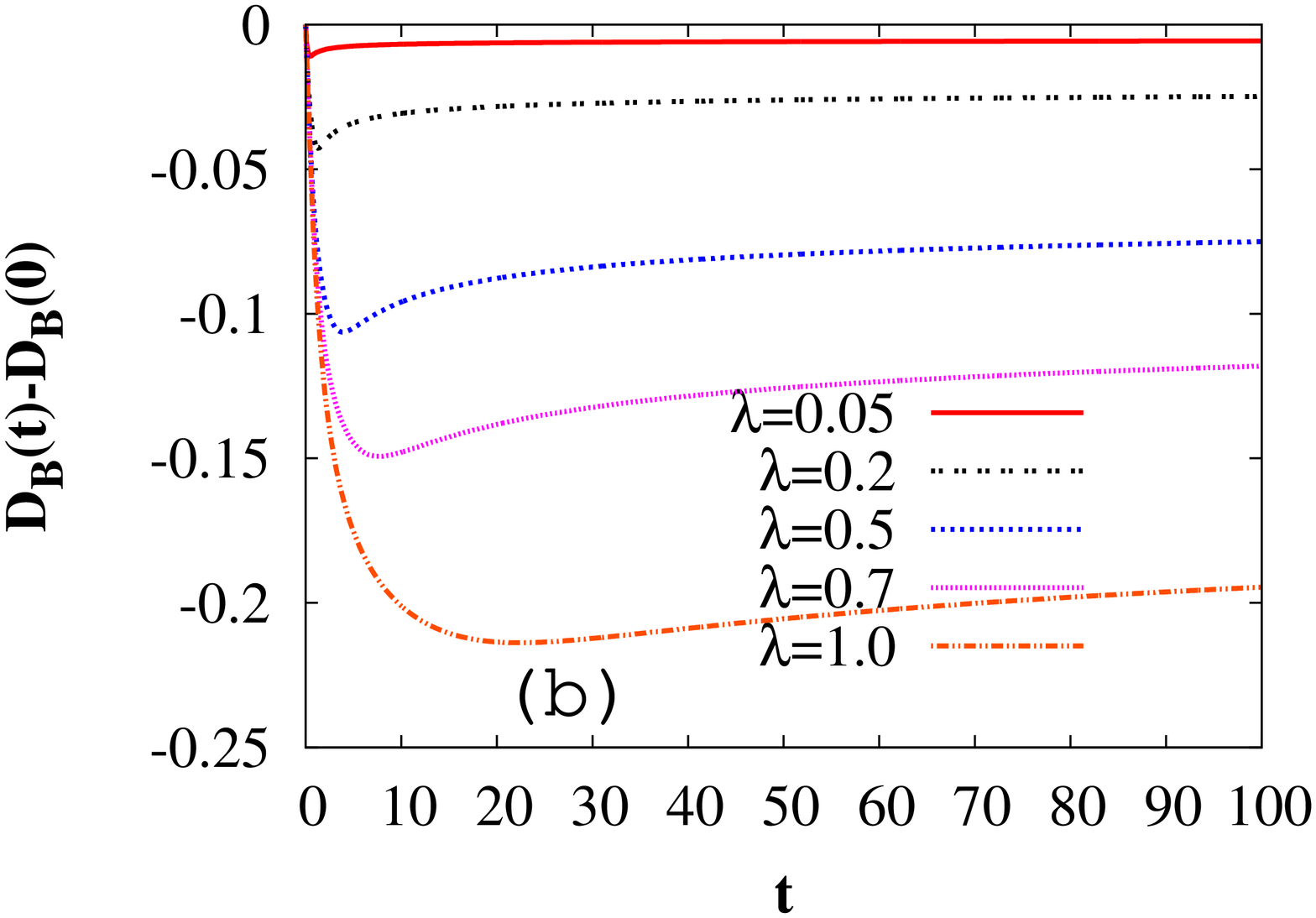}
\includegraphics[width=0.2\textwidth, angle=270]{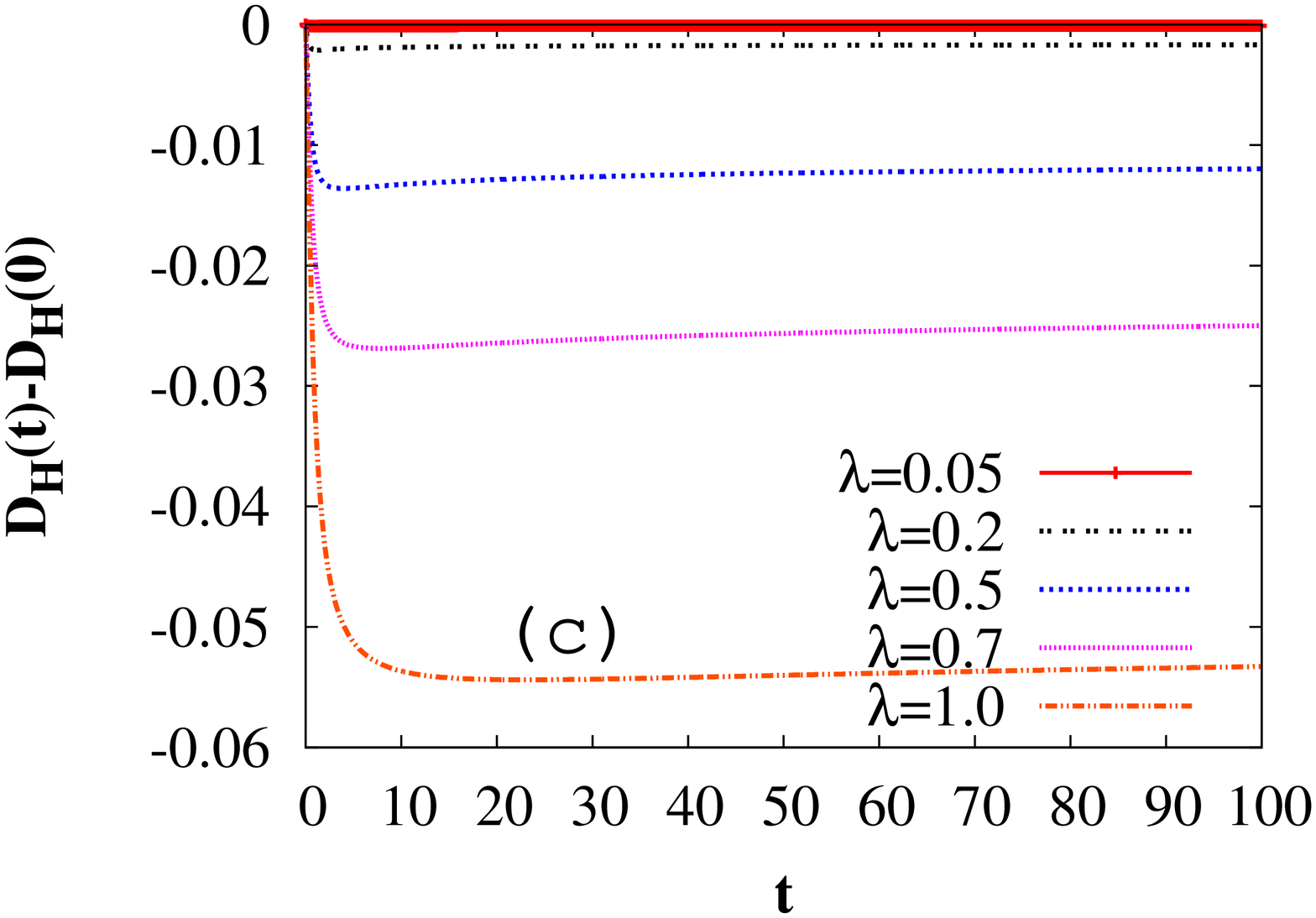}
\includegraphics[width=0.2\textwidth, angle=270]{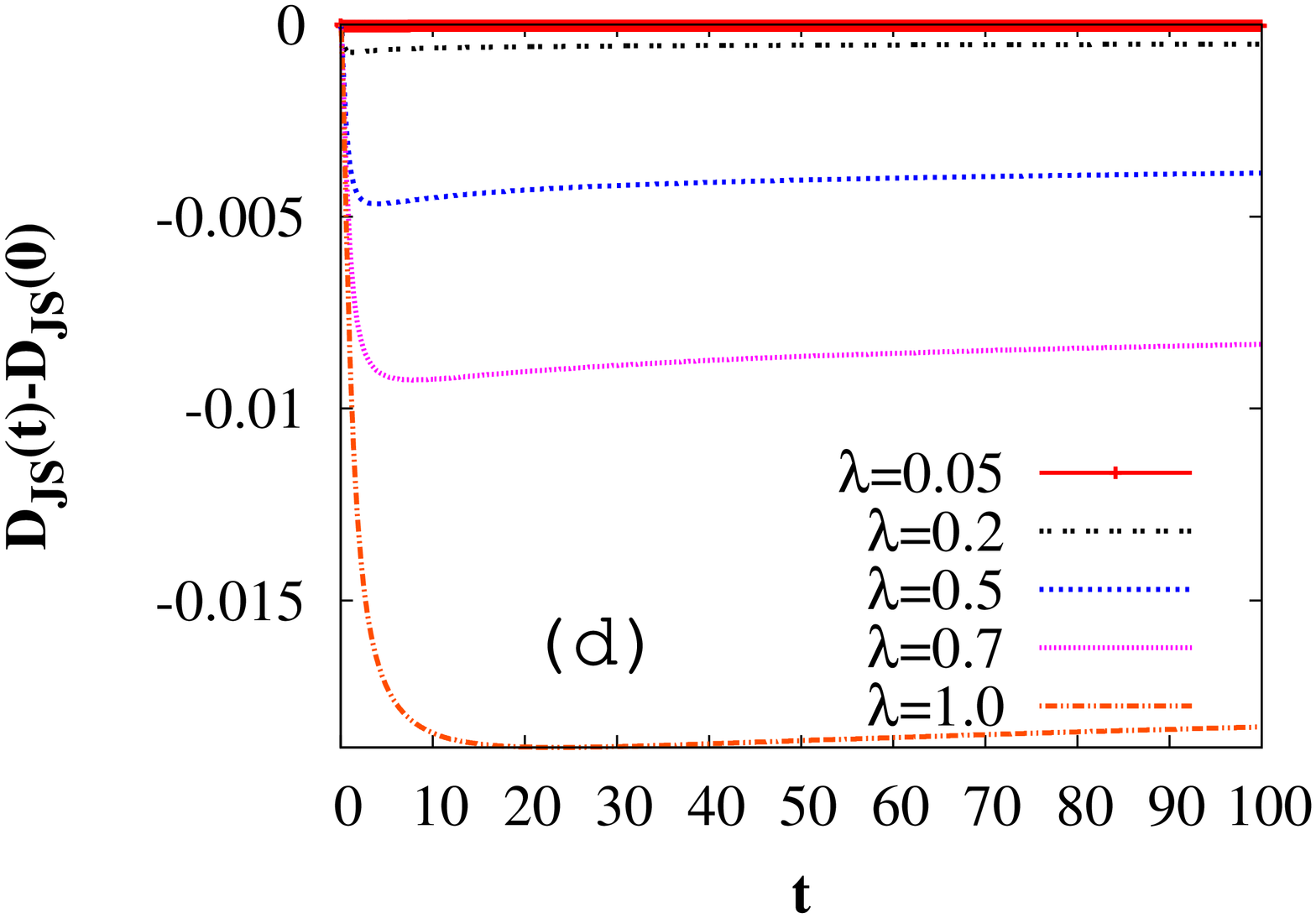}
\caption{(Color online) Time evolution of  distances between two qubit states in the case of infinite environment. Panel (a): the  trace $D_T = D_{HS}/\sqrt{2}$,  panel (b): Bures $D_B$,  panel (c):  Hellinger $D_H$ and panel (d): quantum Jensen-Shannon $D_{JS}$ distances, respectively. The distances  $D(t)=D[\rho_{0}(t), \rho_{\lambda}(t)]$ are  between the initially  non-correlated  and correlated  states for selected values of the correlation parameter $\lambda$.  Time is in unit of $\omega_c$, the dimensionless coupling $\alpha \omega_c^{\mu} = 0.01$ and   $\gamma \omega_c^{\nu}=0.05$.    The remaining parameters are: $\varepsilon =1,   \mu=0.01$, $\nu=0.2$
and $|b_+^{(1)}|^2 = |b_+^{(2)}|^2 =1/2$.
}
\label{fig1}
\end{figure}
%

\subsection{Analysis of different distance measures}

For the analysis of distance properties of the model considered, we still have to specify two quantities: the spectral density $g_h(\omega) = g(\omega)/h(\omega)$  and the coherent state determined by the function $f(\omega)$.
The spectral density function $g_h(\omega)$
completely defines the coupling and modes of the environment.
Typically the spectral function is taken as some continuous
function of frequency to indicate that the environment can be treated as infinite compared to the system. With this study we restrict ourselves to the case in which this function assumes the explicit form
\begin{eqnarray}\label{J}
g_h^2(\omega)=\alpha  \, \omega^{\mu-1}\exp(-\omega/\omega_c),
\end{eqnarray}
where $\alpha >0$ is the qubit-environment coupling constant, $\omega_c$ is a cut-off  frequency and $\mu>-1$ is the "ohmicity" parameter:
the case $-1<\mu <0$  corresponds to the sub--ohmic, $\mu =0$ to the ohmic and $\mu >0$  to super--ohmic environments, respectively.  
Comparing this equation with the expression for the standard spectral function $J(\omega)$ 
(see e.g. Refs. \cite{leg,petruc}), one can find the relation   \cite{fidel} 
\begin{eqnarray}\label{gJ}
J(\omega)  =  {\omega^2} g_h^2(\omega).     
\end{eqnarray}
As follows from our previous study, only in the case of super--ohmic environment, the trace distance can increase. Therefore below we analyze only this regime.

\noindent To determine the coherent state $|\Omega_f\rangle$, we can propose any integrable function $f(\omega)$ but for convenience  let
\begin{eqnarray}\label{f}
f^2(\omega)&=&\gamma \, \omega^{\nu-1}\exp(-\omega/\omega_c).
\end{eqnarray}
The only reason for such a choice is possibility to calculate explicit formulas
 for  the functions in Eqs. (\ref{r}) and (\ref{Fi}). As a result one gets

\begin{eqnarray}\label{LL}
r(t)=4\mathcal{L}(\alpha,\mu,t)],   \quad \quad  \quad     \nonumber \\
s(t)= 2\mathcal{L}(\sqrt{\alpha \gamma},(\mu+\nu)/2,t)-\frac{1}{2} \gamma\Gamma(\nu)\omega_c^\nu,
  \nonumber \\
\ \ \ \nonumber \\
\Phi(t) = \sqrt{\alpha \gamma}\; \Gamma\left(\frac{\mu+\nu}{2}\right)
 \omega_c^{\frac{\mu+\nu}{2}}
\;  \frac{\sin\left[\frac{\mu+\nu}{2} \arctan(\omega_c t) \right] }{(1+\omega_c^2t^2)^{\kappa/2}}, \nonumber
\end{eqnarray}
%
%
\begin{eqnarray}\label{el}
\mathcal{L}(\alpha,\mu,t) =\alpha\Gamma(\mu) \omega_c^\mu\left\{1-\frac{\cos\left[\mu\arctan(\omega_c t) \right] }{(1+\omega_c^2t^2)^{\mu/2}}\right\}
\end{eqnarray}
and $\Gamma(z)$ is the Euler gamma function.

We next    examine  the time evolution of the distance for all four distance measures: namely
the  trace distance $D_T$, the Bures distance $D_B$, the Hellinger distance $D_H$  and the quantum Jensen-Shannon measure $D_{JS}$.
We recall that the trace and  Hilbert-Schmidt distances are equivalent.  As shown in Ref. \cite{daj}, the only chance  to observe  an increase of the distance between two states is  to vary the parameters of environment encoded in $|\Omega_\lambda\rangle$ in Eq.(\ref{omega1}). The simplest theoretical possibility is to manipulate the {\it correlation parameter} $\lambda$. When two different states are determined by two different sets of numbers $b_\pm^{(k)} \;(k=1, 2)$ in Eq.  (\ref{ini})  for the same state $|\Omega_\lambda\rangle$
then $ A_{\lambda_{1}}(t)  =  A_{\lambda_{2}}(t)$ and an increasing growth of the distance becomes not possible.

In Fig. \ref{fig1}, we depict the  time evolution of the  distances $D(t)=D[\rho_{0}(t), \rho_{\lambda}(t)]$   between the initially  non-correlated  and correlated  states for four metrics. We observe that only for the trace metric, the distance $D[\rho_{0}(t), \rho_{\lambda}(t)]$ can increase above its initial value and  there is some optimal value of the correlation parameter  $0 < \lambda < 1$  for which the   distinguishability  of  final  states is the best. Because this case was studied in Ref. \cite{daj}, we do not present the  details  here for the trace distance properties.
  In the remaining three cases, the distance between  states at arbitrary time $t>0$   is always smaller  than the distance at time $t=0$ and the distinguishability of final states is weaker than for the initial states.     An interesting feature  is the appearance of the absolute minimal distance at some time $t_m >0$ during the time evolution of the qubit. At  an early stage of time evolution, the distance decreases, reaching a minimum before it   increases again and eventually saturates  at asymptotic  long times. The conclusion from our analysis depicted in Fig. \ref{fig1} hence is as follows:  An increase of the distance above its initial value between two qubit states presents not a universal property of the correlated initial state but instead is  rather sensitive to the chosen metric measure. Among our chosen five different  metric measures, only  the trace and the Hilbert-Schmidt  metrics exhibit this typical property for the considered decoherence model.
\begin{figure}[t]
\begin{center}
\includegraphics[width=0.2\textwidth, angle=270]{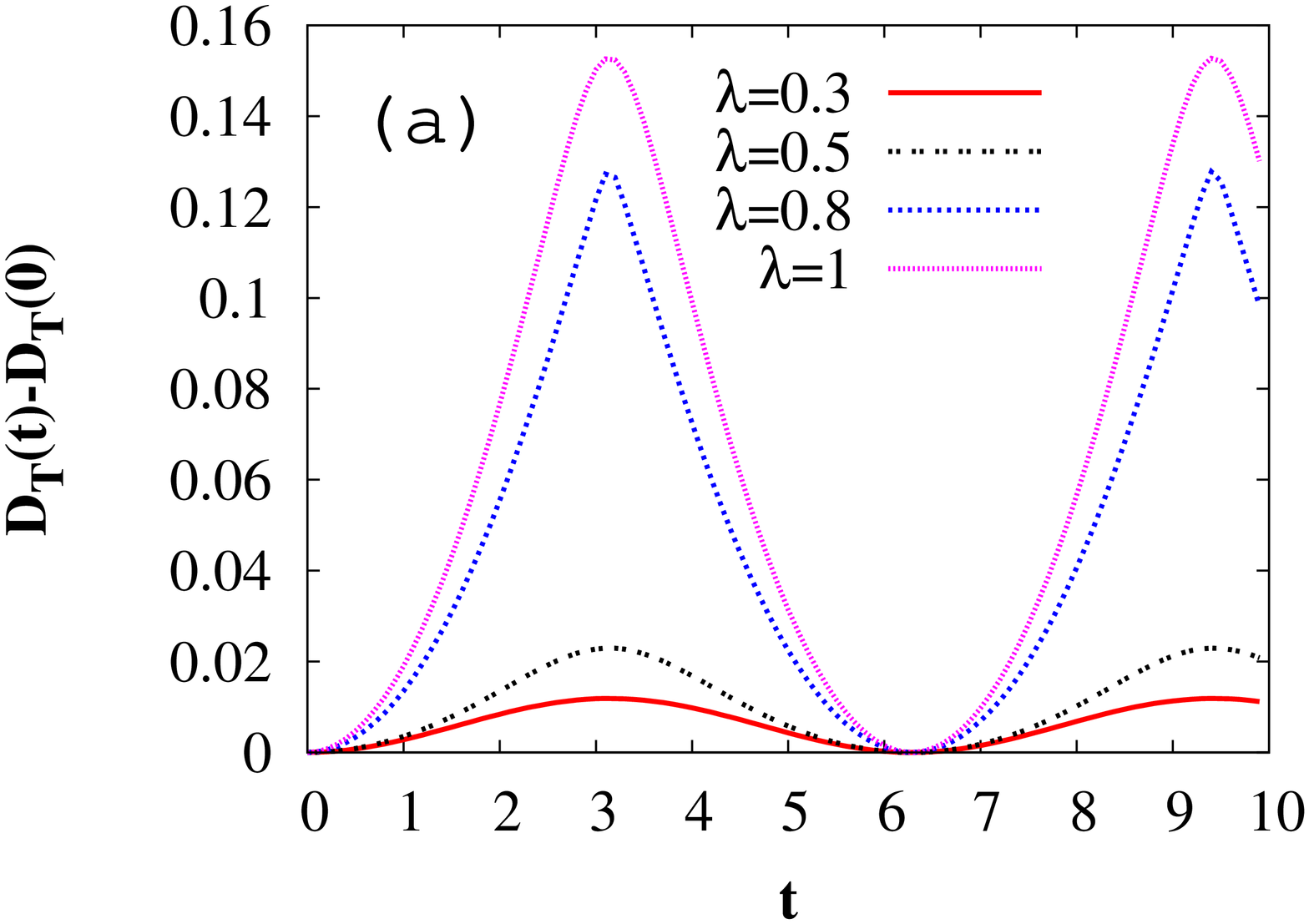}
\includegraphics[width=0.2\textwidth, angle=270]{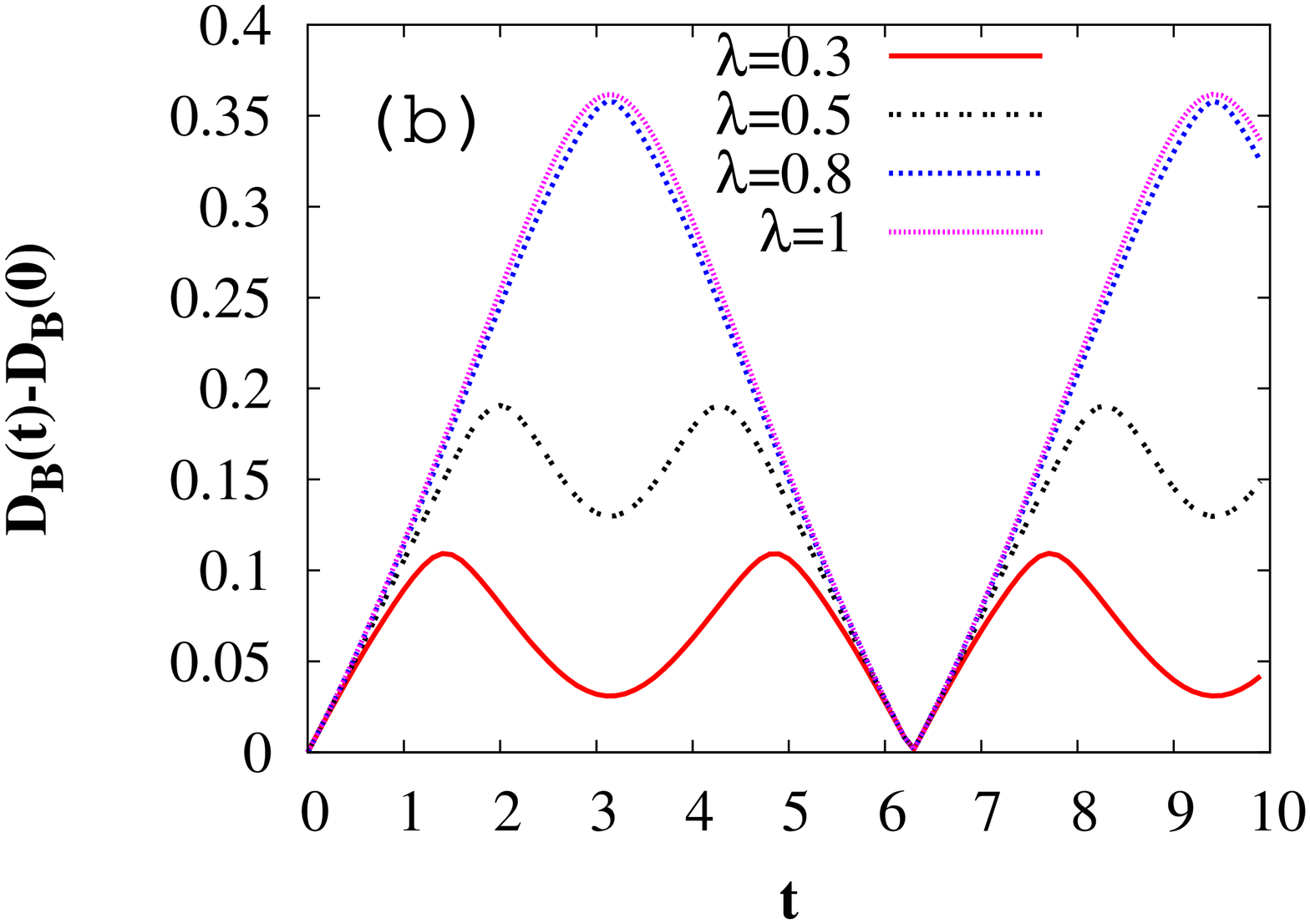}
\includegraphics[width=0.20\textwidth,angle=270]{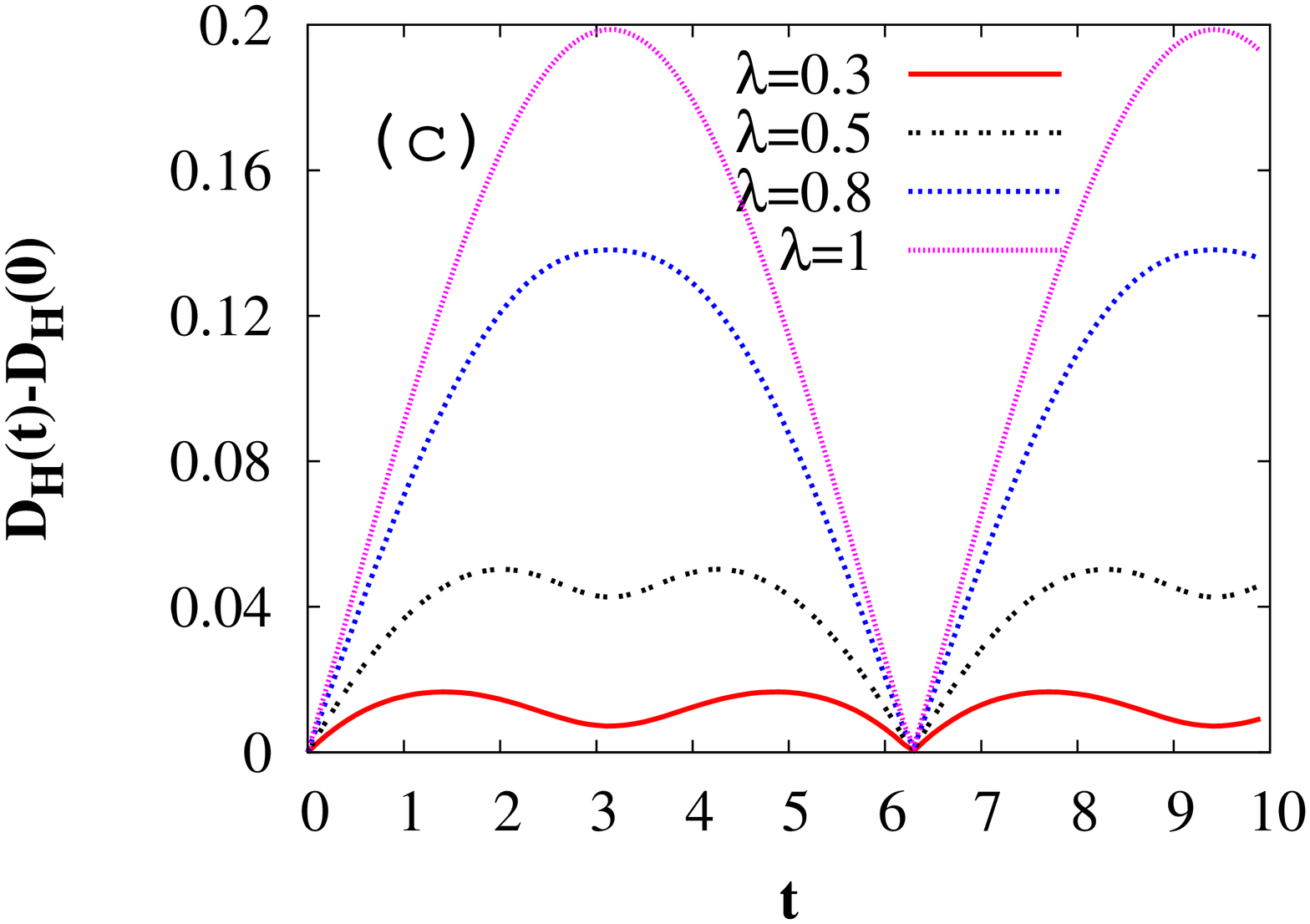}
\includegraphics[width=0.2\textwidth, angle=270]{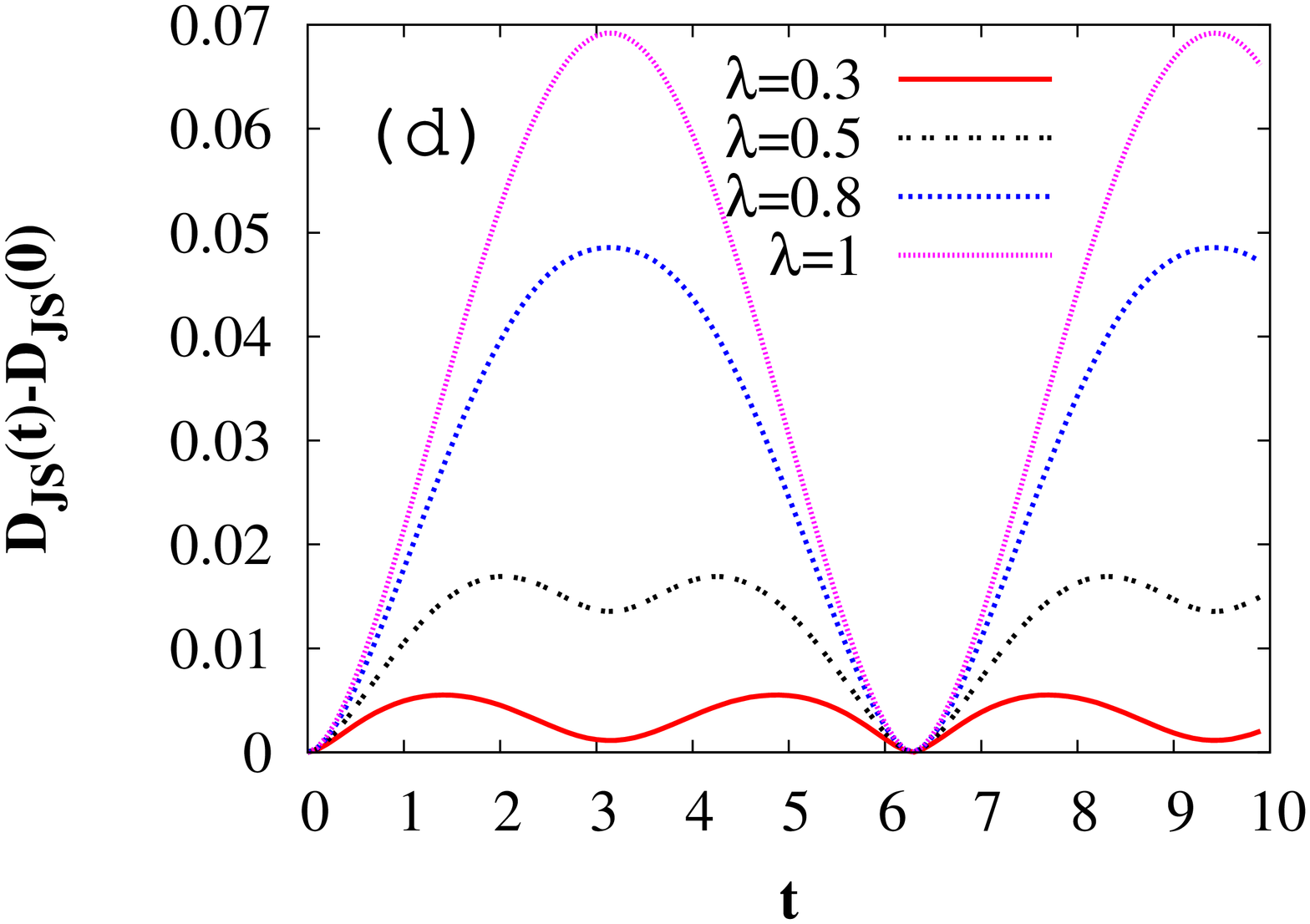}
\end{center}
\caption{(Color online) Time evolution of the (a): trace $D_T = D_{HS}/\sqrt{2}$, (b): Bures $D_B$, (c): Hellinger $D_H$  and (d):  Jensen-Shannon $D_{JS}$ distances  of qubit states for finite environment.  Initially the boson is  in the   mixture of the ground $|0\rangle$ and coherent state
$|z\rangle$ with  $z=|z|e^{i\phi}$.  The impact of initial correlations quantified  by the parameter $\lambda$ is depicted.  As in Fig. 1, the distances  $D(t)=D[\rho_{0}(t), \rho_{\lambda}(t)]$ are  between the initially  non-correlated  and correlated  states. Time is in unit of $\omega$. The chosen system parameters are: $\varepsilon =1, g=0.1, |z|=1, \phi =0$  and $|b_+^{(1)}|^2 = |b_+^{(2)}|^2 =1/2$.   }
\label{fig2}
\end{figure}
%
%


\section{Model B: Qubit coupled to a finite environment}

The preparation of an initial state as determined by Eq.  (\ref{ini})  requires highly sophisticated quantum engineering tools which  presently seem not feasible or at best difficult to realize. Fortunately,  interesting features of distances between states resulting from initial system--environment correlations can be studied with a simplified setup.
Following such reasoning we next   study a qubit that is coupled to {\it finite} size environment. In this case the notion of   decoherence is absent  in a strict sense of the term. Nevertheless, the considered qubit constitutes an open system. Our  choice of a finite bosonic environment is motivated by  recent progress in quantum engineering of non--classical electromagnetic fields
which can be prepared in various states, both in the optical \cite{opt} and  in the microwave \cite{mv} energetic regimes.
As an example, we consider a  single boson mode. The total Hamiltonian (\ref{H1}) then reduces to the form
\begin{eqnarray} \label{H2}
%
%
%
H=diag[H_{+}, H_{-}],  \quad \nonumber\\
H_\pm = \omega a^\dagger a \pm g_0(a+a^\dagger)\pm\varepsilon { \mathbb{I}}_B,
\end{eqnarray}
where $g_0$ is a coupling constant.
The initial state of the total  system is in general correlated, namely,
\begin{eqnarray}\label{fin_ini}
|\Psi(0)\rangle=b_+|1\rangle\otimes |0\rangle+b_-|-1\rangle\otimes |\Omega_\lambda\rangle,
\end{eqnarray}
where
\begin{eqnarray}
|\Omega_\lambda\rangle=C_\lambda^{-1} \left[(1-\lambda)|0\rangle+\lambda|F\rangle \right].
\end{eqnarray}
The state $|0\rangle$ is a vacuum state (a ground state) of the boson  and the choice for the  state
$|F\rangle$ is limited to two classes studied in quantum optics, being known to be distinct with respect to their non--classical character. First we use  $|F\rangle=|z\rangle$ to be a coherent state. Next we analyze the case when $|F\rangle=|N\rangle$ is a number eigenstate.   The density matrix of the qubit assumes the same structure as in Eq. (\ref{ro}), but now with the  modified function $A_{\lambda}(t)$.

%
\begin{figure}[t]
\begin{center}
\includegraphics[width=0.2\textwidth, angle=270]{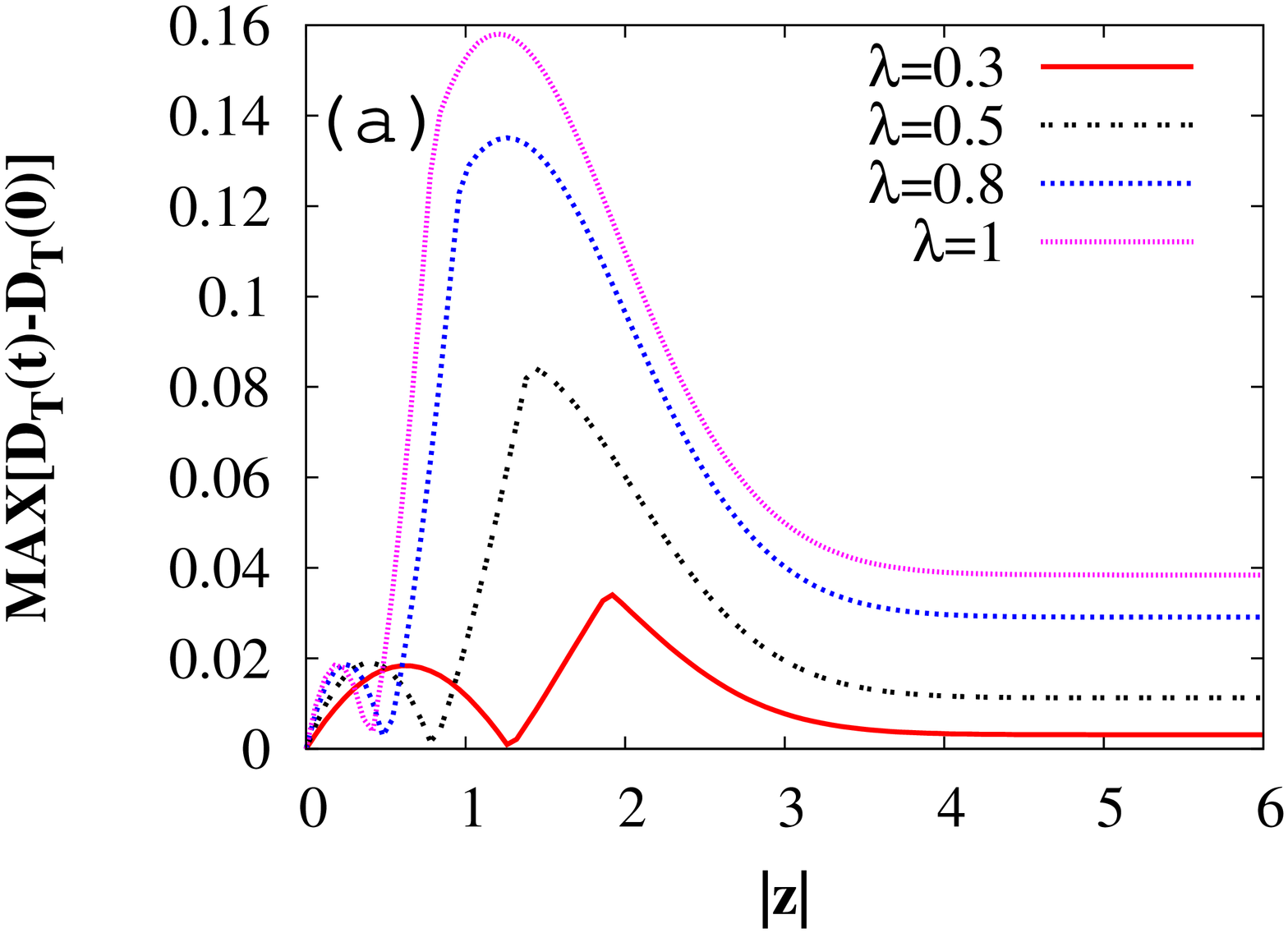}
\includegraphics[width=0.2\textwidth, angle=270]{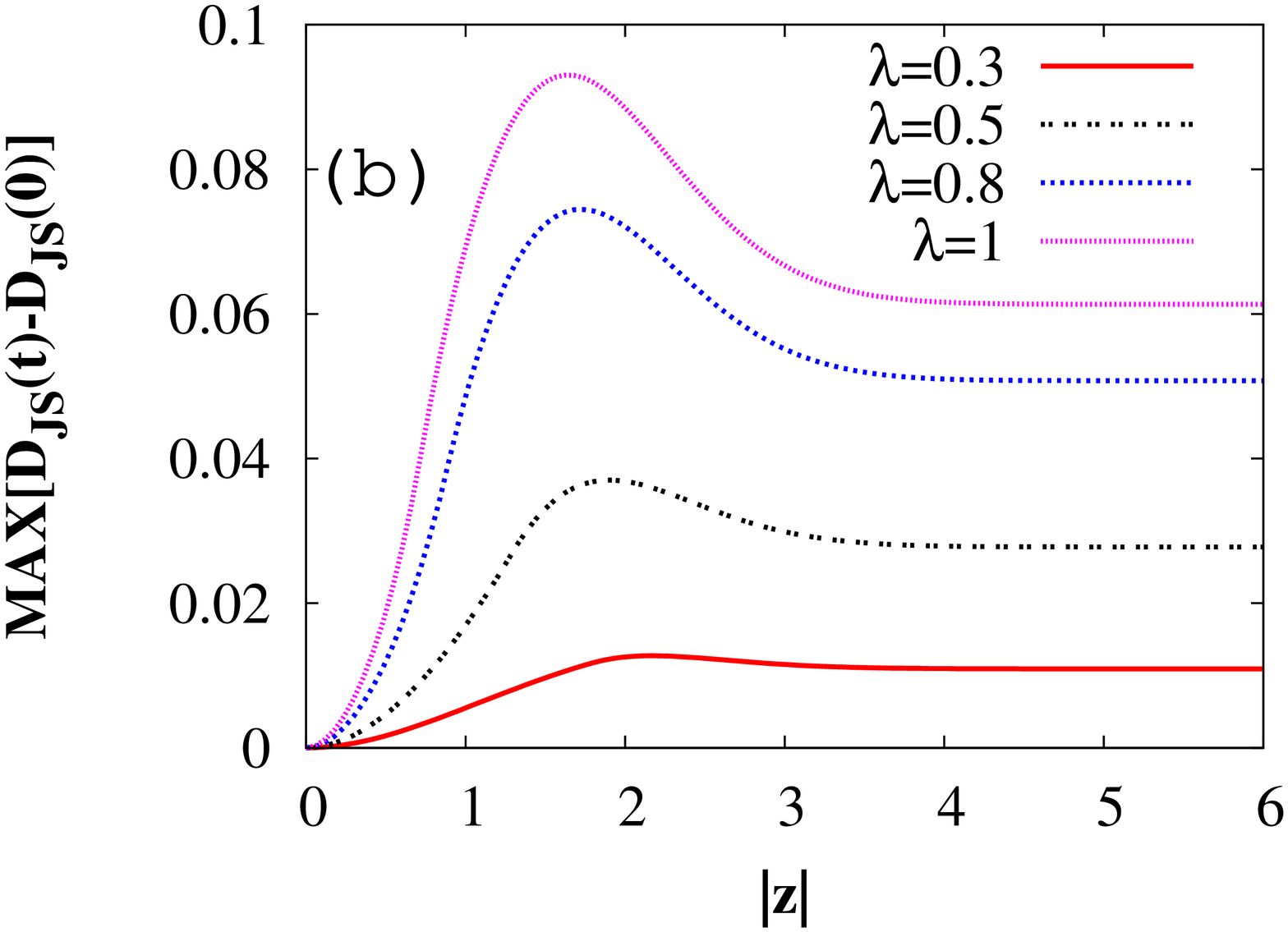}
\includegraphics[width=0.2\textwidth, angle=270]{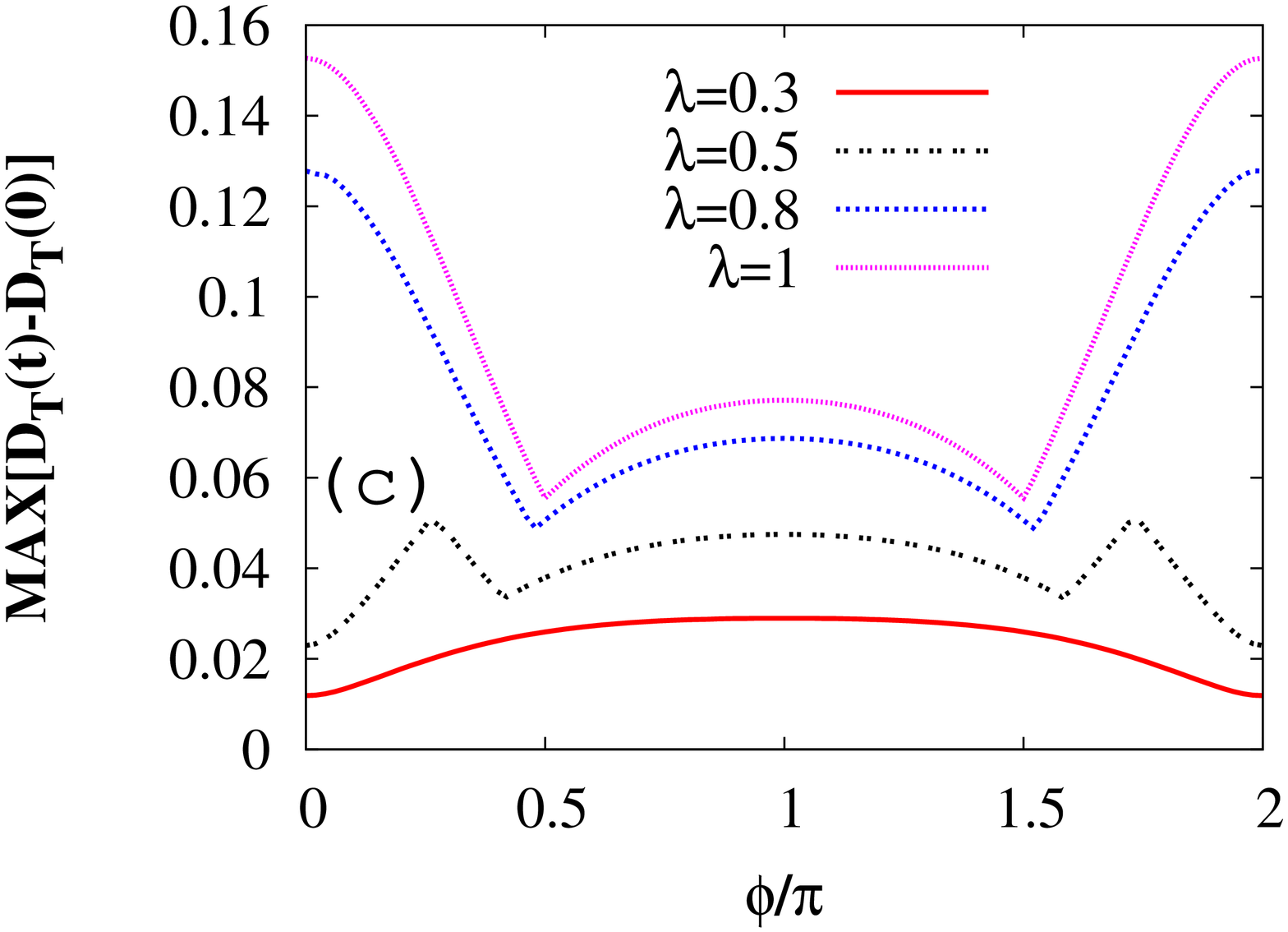}
\includegraphics[width=0.20\textwidth,angle=270]{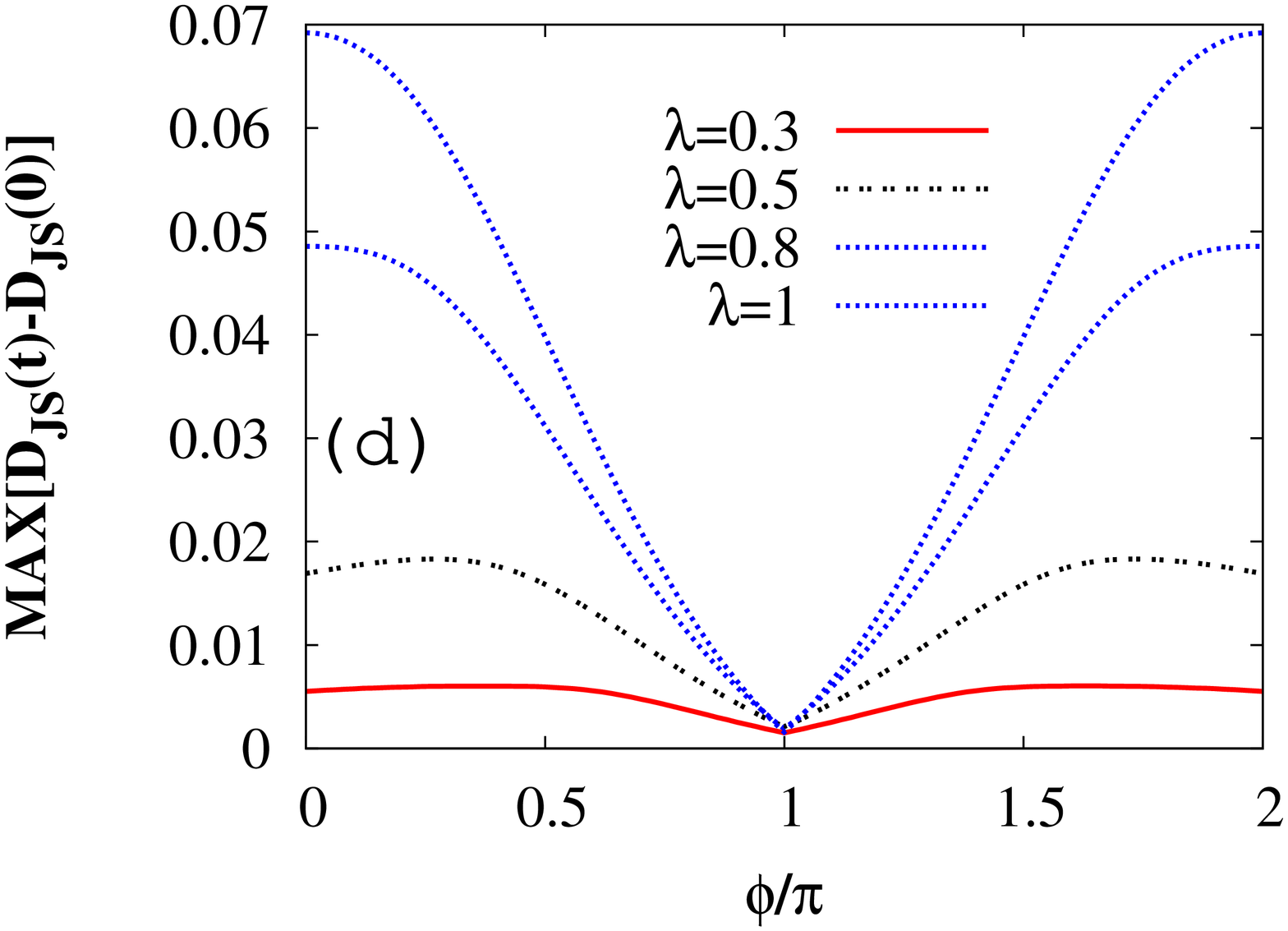}
\end{center}
\caption{(Color online)  Illustration of role of the amplitude $|z|$ (panels (a) and (b)) and phase $\phi$  (panels (c) and (d)) of the  environment coherent state $|z\rangle$ ($z=|z|e^{i\phi}$)  on trace and Jensen-Shannon distances  of qubit states. Qualitatively, the Bures and Hellinger distances behave like  Jensen-Shannon distance.
MAX$[D(t)-D(0)]$ is the amplitude of distance time oscillations  shown in Fig. 2. In panels (a) and (b):
$\phi=0$. In panels (c) and (d): $|z|=1$.
 The remaining parameters are  the same as in Fig. \ref{fig2}.}
\label{fig3}
\end{figure}

\subsection{The case of  initial coherent states}

Let for any complex number $z=|z|e^{i\phi}$,  the state $|F\rangle=|z\rangle$
be a coherent state of the boson. Then the function $A_{\lambda}(t)$ is given by
\begin{eqnarray}\label{Acoh}
 A_{\lambda}(t)= C_{\lambda}^{-1} \,e^{-2i\varepsilon t - R(t)} \, \left[1-\lambda+\lambda
e^{-2i\Lambda(t) + S(t)}  \right], \quad
\end{eqnarray}
where
\begin{eqnarray}   \label{a0}
R(t)&=&4g^2 [1-\cos(\omega t)], \nonumber \\
S(t) &=& 2g|z|[\cos\phi - \cos(\omega t -\phi)] - \frac{1}{2} |z|^2,  \nonumber \\
\Lambda(t) &=& g|z| [ \sin(\omega t +\phi)] + \sin \phi]
\end{eqnarray}
and  $g=g_0/\omega$ is rescaled coupling.

\begin{figure}[t]
\begin{center}
\includegraphics[width=0.2\textwidth, angle=270]{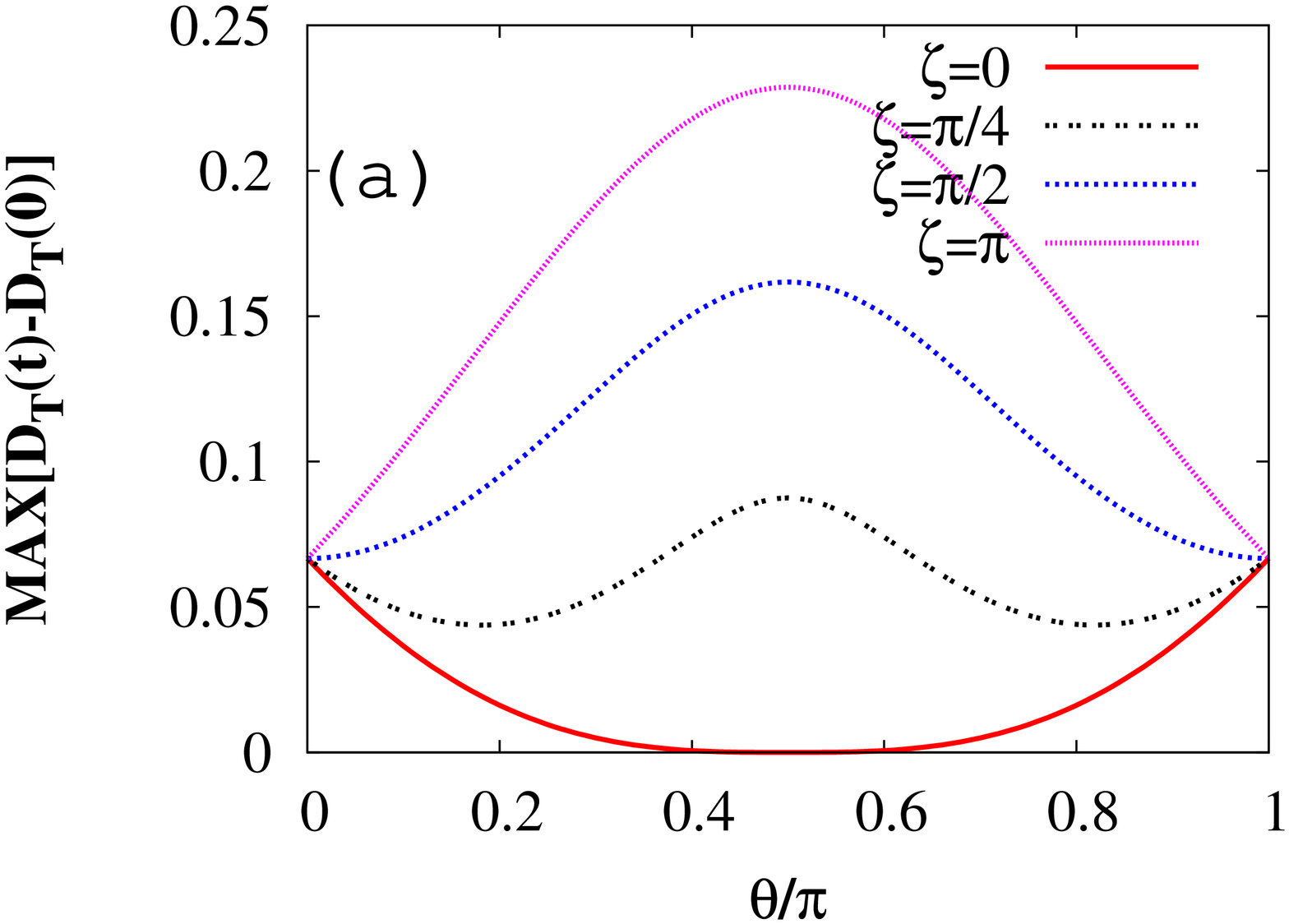}
\includegraphics[width=0.20\textwidth,angle=270]{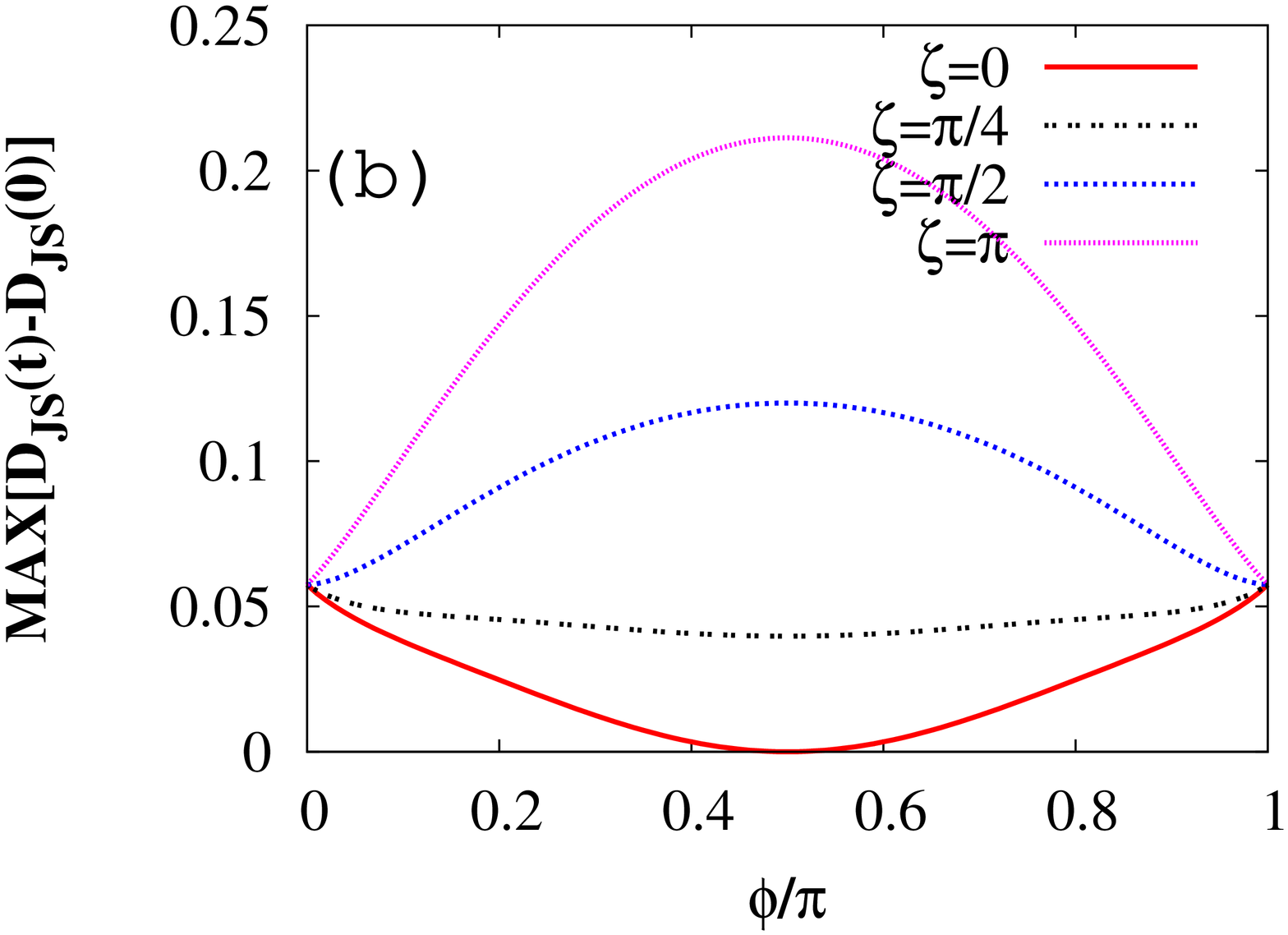}
\end{center}
\caption{(Color online) Illustration of role of initial qubit states on trace and Jensen-Shannon distances.  Qualitatively, the Bures and Hellinger distances behave like  Jensen-Shannon distance. MAX$[D(t)-D(0)]$ is the amplitude of distance time oscillations  shown in Fig. 2. The parameters characterizing  two  initial states (\ref{ini}) are: $\lambda_1 =\lambda_2 =1$, for the first state:  $b_+=b_-=1/\sqrt{2}$,  for the second state:  $b_+=\cos(\theta/2),   b_-=\exp( i \zeta) \sin(\theta/2)$ with  angle parameterization  $\theta$ and $\zeta$ on the Bloch sphere.
The remaining parameters are: $g=0.1, z=1$ }
\label{fig4}
\end{figure}
Because the total system is finite,  time evolution of the qubit states is periodic.   However, it is not unitary evolution.
The distance between two states of the qubit is also a periodic function of time. Let us now inspect time-dependence of all four distance: trace $D_T = D_{HS}/\sqrt{2}$, Bures $D_B$, Hellinger $D_H$  and  Jensen-Shannon $D_{JS}$ distances. In Fig. \ref{fig2} we illustrate  the role of initial qubit-environment correlations in the case when  two different initial states are determined by two different states $|\Omega_\lambda\rangle$ with different  $\lambda_1$ and $\lambda_2$.  The most peculiar feature is that for all measures, the distance at any time $t>0$ is not smaller than at initial time. It is in clear contrast to the case of infinite environment case when only the trace distance can increase about its initial value.  Now,  at the beginning, for $t>0$, all distances increase above its initial value reaching the maximal value which in turn grows when the correlation parameter $\lambda \to 1$.  The maximal amplitude of distance oscillations  is shown up for  maximally entangled states, i.e. for $\lambda = 1$.
It also depends on other system parameters, in particular on the state of environment which is determined by two quantities: the amplitude $|z|$ and phase $\phi$ of the coherent state $|z\rangle$.
The inspection of the results revealed that there
are regimes of optimal values of $|z|$ for which distinguishability of two qubit states is most prominent.  We present it in Fig. \ref{fig3} for  the trace and Jensen-Hellinger distances. The remaining two  (Bures and  Hellinger) distances  exhibit similar behavior like  the Jensen-Shannon distance. In two bottom panels of Fig. \ref{fig3} we demonstrate how the phase of the coherent state changes the distance. Again, as previously, we present only two cases. Two other cases are similar
to the Jensen-Shannon one. Let us observe that in some regimes the trace distance possesses distinctive features which are different from  other distance measures.

Next, let us consider the case when  two different states are determined by two different sets of numbers $b_\pm^{(k)} \;(k=1, 2)$ in Eq.  (\ref{ini}) but  with the same state $|\Omega_\lambda\rangle$. One state is fixed by $b_+=b_-=1/\sqrt{2}$. The second state is conveniently parameterized by two angles $\theta$ and $\zeta$ on the Bloch sphere and is determined by relations: $b_+=\cos(\theta/2)$ and $b_-=\exp( i \zeta) \sin(\theta/2)$.  The result is depicted in Fig. \ref{fig4} which shows that the
amplitude of time-periodic oscillations of the distance can typically  be increased by increasing the geometrical distance  of initial states on the Bloch sphere. However, there are some exceptions such as, for example, for the case $\zeta=\pi/4$ in the case of the trace distance.

\begin{figure}[t]
\begin{center}
\includegraphics[width=0.2\textwidth, angle=270]{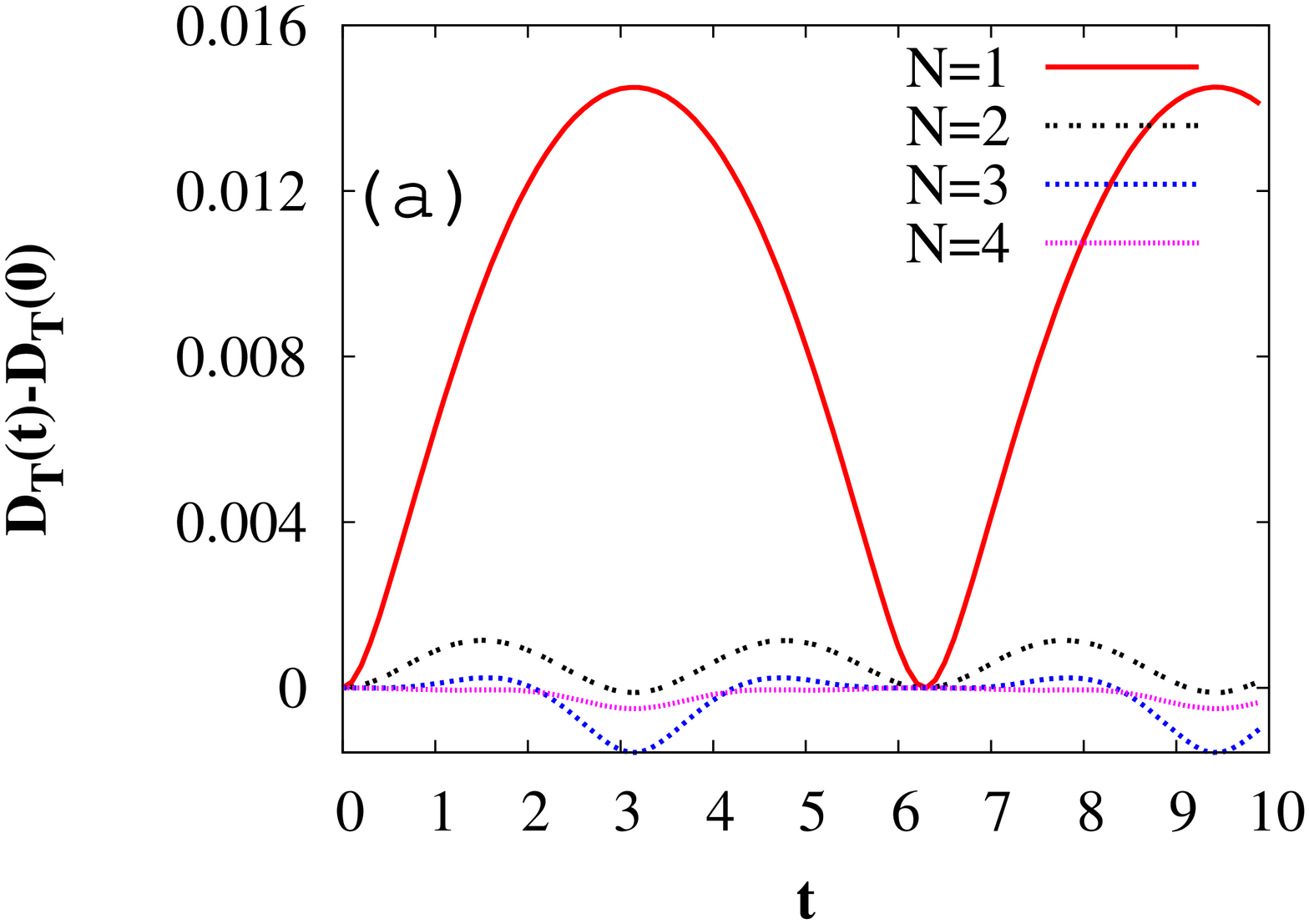}
\includegraphics[width=0.20\textwidth,angle=270]{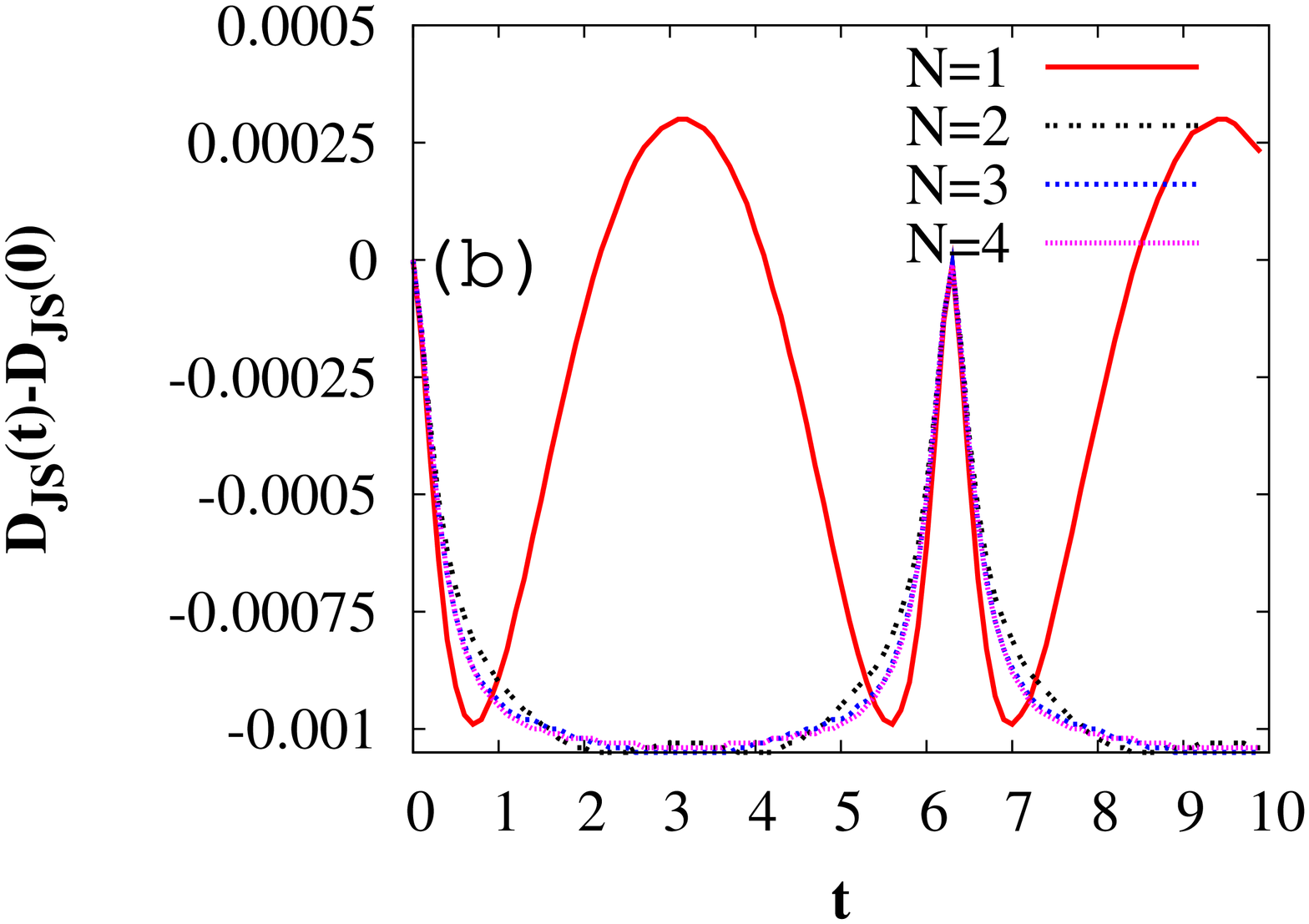}
\end{center}
\caption{(Color online) The trace and Jensen-Shannon distances  of qubit states for finite environment: boson in the   mixture of the ground $|0\rangle$ and number $|N\rangle$ states. The Bures and Hellinger distances display similar time-dependence as the Jensen-Shanon distance with the exception that they lie below zero.  The parameters are: $\lambda_1=0, \lambda_2=1, \epsilon=1, g=0.1$
and $|b_+^{(1)}|^2 = |b_+^{(2)}|^2 =1/2$. }
\label{fig5}
\end{figure}
%


\subsection{The case of initial number states}
Let $|F\rangle=|N\rangle$ be   a number eigenstate of the boson. Contrary to  coherent states,
such eigenstates are orthogonal and the state (\ref{fin_ini}) becomes maximally entangled, i.e., its partial trace, taken with respect to the bosonic degree of freedom,  is an identity and it corresponds to the  maximally mixed state of the  qubit.
In this case, the function $A_{\lambda}(t)$ assumes the form
\begin{eqnarray}\label{Acoh1}
 A_{\lambda}(t)= C_{\lambda}^{-1} \,e^{-2i\varepsilon t - R(t)} \, \left[1-\lambda+\lambda
B_N(t) \right],
\end{eqnarray}
where
\begin{eqnarray} \label{B}
B_N(t)=\frac{(2g)^N}{\sqrt{N!}} \left(e^{-i\omega t} -1\right)^N.
\label{a1}
\end{eqnarray}
As in the former case,  time-evolution of the qubit states is time-periodic and in consequence distance is also periodic function of time.
In Fig. \ref{fig5},  we present two forms of the distance, namely the trace and the Jensen-Shannon ones.
Only these two distance measures can exhibit the increase distance over its initial value. 

The 'optimal' environment state is the first excited one, i.e., when $N=1$.
 This state is highly non--classical. The question whether there is any relation between non--classical character of the environment and the distance between reduced qubit states remains open and will be postponed for further considerations.  Further excited states diminish the positive value of difference  $D(t)-D(0)$ or invert it into negative value.    Two remaining (Bures and Hellinger) distances behave in a similar way as the Jensen-Shannon one but they are removed down and never  exceed their initial values.

%

\section{Summary}

The objective to distinguish two quantum channels presents a most important challenge for quantum
information processing tasks.
The difficulty of the distinguishability issue leads naturally to a study of the
problem on restricted classes of channels. With this work we  presented two models 
and  we have elucidated the properties of four distance measures for quantum  states for the situation of a qubit which is coupled to an environment. At initial times, the system is in a correlated (entangled) state.  Our chosen measures include  the trace (and equivalent Hilbert-Schmidt), Bures,
Hellinger and Jensen-Shannon distances. We have considered two examples of the environment: namely an infinite one consisting of bosons and finite one consisting of a single boson. We have demonstrated that in the case of the infinite environment, only the trace distance exhibits an increase above its initial value. All other remaining distances studied do not exhibit this property. In the case of a finite environment, however,
some kind of universality is observed for the case when the boson consists  in a mixture of  ground and coherent states. In this latter case, all distances behave  more or less similarly: the distance measures oscillate with a common frequency
between an  initial value and some maximal positive value, which is different for differently chosen metrics.
Nevertheless, their time dependence  behaves  qualitatively  the same. This is not the case when
the boson is in a mixture of the ground and excited states; only the trace  and Jensen-Shannon distances
are allowed to grow above the initial value.

Our  main conclusion is as follows:  the result of an increase of the distance measure  above its initial value constitutes  no universal property; its behavior upon evolving time  strongly  depends on the employed distance measure; in this respect, the trace distance receives a special status.

We authors are confident that this work  may  stimulate yet additional studies. Particularly,
it would be interesting to investigate in some detail the objective of universally valid, initial-state dependent
and/or system-dependent properties of the various distance measures in use. Generalization of our results to (i)  other classes of initial correlations between the system and  environments of different nature and (ii) for non-zero temperatures  provide yet other appealing routes for  future research.\\
\\

\section*{Acknowledgment}
The work  supported by   the  MNiSW Grant N202 052940 (J.D. and J. {\L}) and the DFG via the collaborative research centre SFB-631 (P.H.), project A5, and the ``Nanosystems Initiative Munich'' (NIM) (P.H.). We authors acknowledge as well the support by the ESF program ``Exploring the Physics of Small Devices".

\end{document}